# Ferromagnetism and the formation of interlayer As$_2$-dimers in Ca(Fe$_{1-x}$Ni$_x$)$_2$As$_2$

Roman Pobel, Rainer Frankovsky, and Dirk Johrendt

Department Chemie der Ludwig-Maximilians-Universität München,

Butenandtstr. 5-13, D-81377 München, Germany

The compounds Ca(Fe$_{1-x}$Ni$_x$)$_2$As$_2$ with the tetragonal ThCr$_2$Si$_2$-type structure (space group *I*4/*mmm*) show a continuous transition of the interlayer As-As distances from a non-bonding state in CaFe$_2$As$_2$ ($d_{As-As}$ = 313 pm) to single-bonded As$_2$-dimers in CaNi$_2$As$_2$ ($d_{As-As}$ = 260 pm). Magnetic measurements reveal weak ferromagnetism which develops near the composition Ca(Fe$_{0.5}$Ni$_{0.5}$)$_2$As$_2$, while the compounds with lower and higher nickel concentrations both are Pauli-paramagnetic. DFT band structure calculations reveal that the As$_2$-dimer formation is a consequence of weaker metal-metal in *M*As$_{4/4}$-layers (*M* = Fe$_{1-x}$Ni$_x$) of Ni-richer compounds, and depends not on depopulation or shift of As-As σ* antibonding states as suggested earlier. Our results also indicate that the ferromagnetism of Ca(Fe$_{0.5}$Ni$_{0.5}$)$_2$As$_2$ and related compounds like SrCo$_2$(Ge$_{0.5}$P$_{0.5}$)$_2$ is probably not induced by dimer breaking as recently suggested, but arises from the high density of states generated by the transition metal 3*d* bands near the Fermi level without contribution of the dimers.

**Introduction**

The discovery of high-$T_c$ superconductivity in iron arsenides has renewed the interest in transition metal pnictides [1-4], and especially compounds with the tetragonal $ThCr_2Si_2$-type structure like $BaFe_2As_2$ and its derivatives rank among the most intensively investigated materials [5-8]. Currently also further structure-property relationships of the $ThCr_2Si_2$-type compounds attract considerable attention, among them the remarkable flexibility of the homonuclear bond between the pnictide atoms of adjacent layers. The whole range from long non-bonding distances up to short single bonded dimers, as well as structural phase transitions between both states have been observed [9-11]. Figure 1 shows the structures of $CaFe_2As_2$ with pairs of isolated $As^{3-}$ ($d_{As-As}$ = 313 pm) [12], and $CaNi_2As_2$ with $As_2^{4-}$ dimers ($d_{As-As}$ = 260 pm) [13]. It is long known that the tendency to form these bonds increases within the $3d$ transition metal period from left to right. A continuous transition between these states has recently been described in solid solutions of the phosphides $Ca(Co_{1-x}Ni_x)_2P_2$ [14].

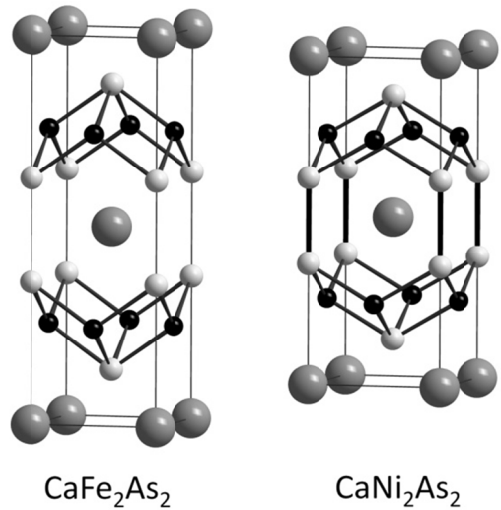

Fig. 1. Crystal structures of $CaFe_2As_2$ and $CaNi_2As_2$

Atom size arguments are obviously ruled out in these cases, thus an electronic origin of these remarkable bond length changes is expected. A widely accepted explanation had been suggested already in 1985 by Hoffmann and Zheng based on semi-empirical band structure calculations of $Mn_2P_2$-layers [15]. They argued that P-P $\sigma^*$ antibonding orbitals become depopulated as the Fermi-level of the metal decreases upon band filling along the $3d$-series. However, subsequent calculations based on density functional theory (DFT) have not supported this concept [16, 17]. Nevertheless, understanding the electronic mechanism of interlayer bond formation is important with respect to the physical properties of $ThCr_2Si_2$-type compounds. The transition from the non-bonded to the bonded state may significantly change the electronic state at the transition metal and also the Fermi surface, both crucial for electronic and magnetic properties. As an example, it is believed that the absence of superconductivity under pressure in $CaFe_2As_2$ [18] in contrast to $BaFe_2As_2$ and $SrFe_2As_2$ [19] has its origin in the formation of $As_2^{4-}$ dimers in $CaFe_2As_2$ at pressures still too low to induce superconductivity, while $BaFe_2As_2$ keeps the structure with non-bonded $As^{3-}$ under pressure. Further examples are magnetic transitions in $EuM_2P_2$ ($M$ = Co, Fe) [20, 21], or the recently observation of ferromagnetism that develops during the course of breaking $Ge_2$-dimers in $SrCo_2(Ge_{1-x}P_x)_2$ [22].

In this article we report the syntheses, crystal structure and magnetism of the solid solution $Ca(Fe_{1-x}Ni_x)_2As_2$ which shows the gradual formation of $As_2^{4-}$ dimers with increasing nickel concentration , while weak ferromagnetism develops near $x \approx 0.5$. Concomitant changes of the electronic structure are studied by DFT band calculations together with bond analysis using the COHP method.

**Results and Discussion**

*Crystal structure*

Figure 2 shows the lattice parameters of the solid solution $Ca(Fe_{1-x}Ni_x)_2As_2$. The *c*-axis decreases strongly with *x* by about −20%, while the *a*-axis slightly increases by 4% only. The changes are not linear, rather *S*-shaped with stronger slopes between $x = 0.3$ and $x = 0.5$. In Fig. 3 we compare the variations of the normalized bond lengths and the twofold angle As-*M*-As ($\varepsilon$). Most strikingly, the As-As distance between the layers becomes shortened by −20%, thus the transition from the non-bonded state in $CaFe_2As_2$ to an $As_2$-dimer in $CaNi_2As_2$ is evident. The enormous shortening is at the expense of increased *M-M* distances (+4%), while the strong *M*-As bonds of the tetrahedra remain almost unaffected. In other words, the $MAs_{4/4}$ tetrahedron becomes flatter, which is manifested in the increasing As-*M*-As bond angle $\varepsilon$.

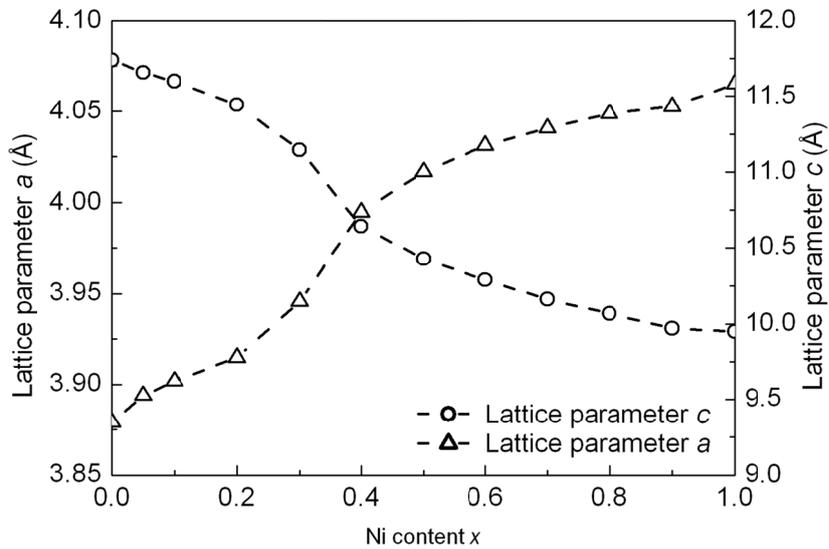

Fig. 2. Lattice parameters of the solid solution $Ca(Fe_{1-x}Ni_x)_2As_2$

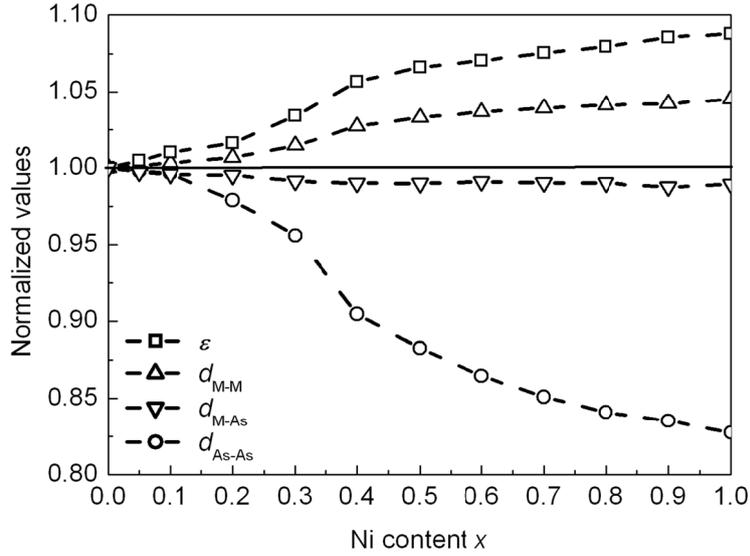

Fig. 3. Normalized changes of bond lengths and the As-$M$-As bond angle ($\varepsilon$) in Ca(Fe$_{1-x}$Ni$_x$)$_2$As$_2$

*Magnetic properties*

The stripe-type antiferromagnetic order of CaFe$_2$As$_2$ ($T_N$ = 173 K) has already been comprehensively investigated [23]. Figure 4 shows the results of magnetic measurements with samples of Ca(Fe$_{1-x}$Ni$_x$)$_2$As$_2$ at $x$ = 0.1, 0.5 and 0.9. The small ($\approx 10^{-4}$ cm$^3$/mol at 150 K) and weakly temperature dependent susceptibilities of Ca(Fe$_{0.9}$Ni$_{0.1}$)$_2$As$_2$ and Ca(Fe$_{0.1}$Fe$_{0.9}$)$_2$As$_2$ are consistent with Pauli-paramagnetic behaviour, while Ca(Fe$_{0.5}$Ni$_{0.5}$)$_2$As$_2$ shows a sudden increase of $\chi$ below 50 K which indicates ferromagnetic ordering. The magnetization isotherm at 300 K is linear, while at 1.8 K (insert of Fig. 4) the magnetization strives after saturation towards a weak moment of 0.2 $\mu_B$/FU. Kink point measurements in field-cooled and zero-field cooled modes (insert of Fig. 4) support the presence of ferromagnetism. The derivatives $d\chi/dT$ yielded Curie temperatures of 47 K (zfc) and 46 K (fc).

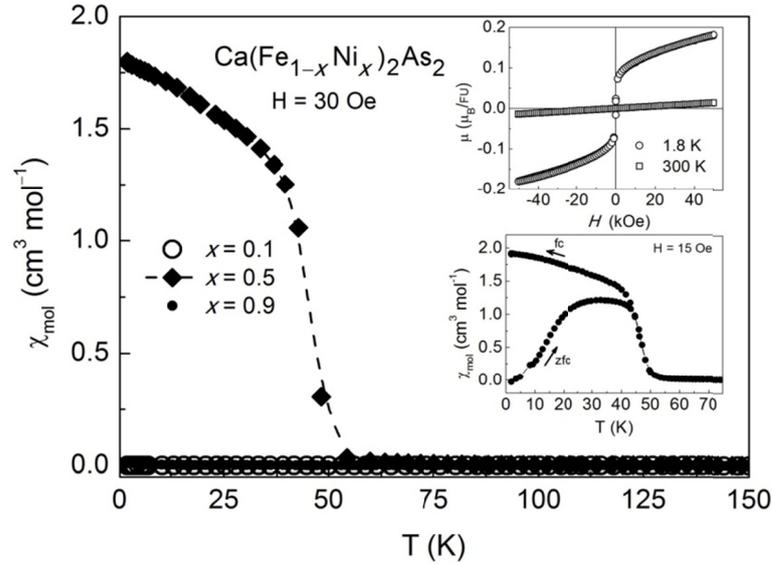

Fig. 4. Magnetic susceptibilities of Ca(Fe$_{1-x}$Ni$_x$)$_2$As$_2$ ($x$ = 0.1, 0.5, 0.9) measured in 30 Oe field. Insert above: Isothermal magnetizations of Ca(Fe$_{0.5}$Ni$_{0.5}$)$_2$As$_2$ at 300 and 1.8 K; below: Low-temperature susceptibility of Ca(Fe$_{0.5}$Ni$_{0.5}$)$_2$As$_2$ measured in zero-field-cooling (zfc) and field-cooling (fc) modes.

*Electronic structure and chemical bonding*

Fig. 5 shows the projected density of states (PDOS) of the transition-metals *M* (solid lines), and the As-contributions (filled areas) in CaFe$_2$As$_2$, CaFeNiAs$_2$ and CaNi$_2$As$_2$, respectively, together with plots of the integrated COHP (ICOHP) of *M*-As, As-As, and *M-M* bonds. ICOHP measures the energy contributions of the specified bonds to the total band structure energy. The PDOS-plots of all three compounds are remarkably similar despite significant changes in the bond lengths and valence electron count (VEC), which increases from 28 in CaFe$_2$As$_2$ to 32 in CaNi$_2$As$_2$. The Fermi level (dashed vertical line) traverses a sharp peak in the PDOS of the metal, and coincides with it in the case of CaFeNiAs$_2$ (VEC = 30). The origin of this peak is the *M-M* antibonding band of $dd\sigma^*$-symmetry. Also remarkable are the similar distributions of the As-PDOS in spite of the large variation of the As-As distances from 313 to 260 pm. In all cases, the As-4$p$ orbitals spread from −6 to +6 eV, while the As-4$s$ orbitals are between −13 and −11 eV. In contrast to the Hoffmann model [15], we find no significant change of the As-band filling in spite of the increased VEC. Fig. 5 also shows that the As-bands contribute sparsely to the density of states at the Fermi level, which is dominated by the metal 3$d$-orbitals. Starting from CaFe$_2$As$_2$, the additional electrons populate mainly 3$d$-bands, and scarcely As-$p$ bands. From this it becomes clear that the observed changes of the crystal structure and physical properties indeed originates from 3$d$-band filling, but not from depopulation of As-As σ* antibonding orbitals.

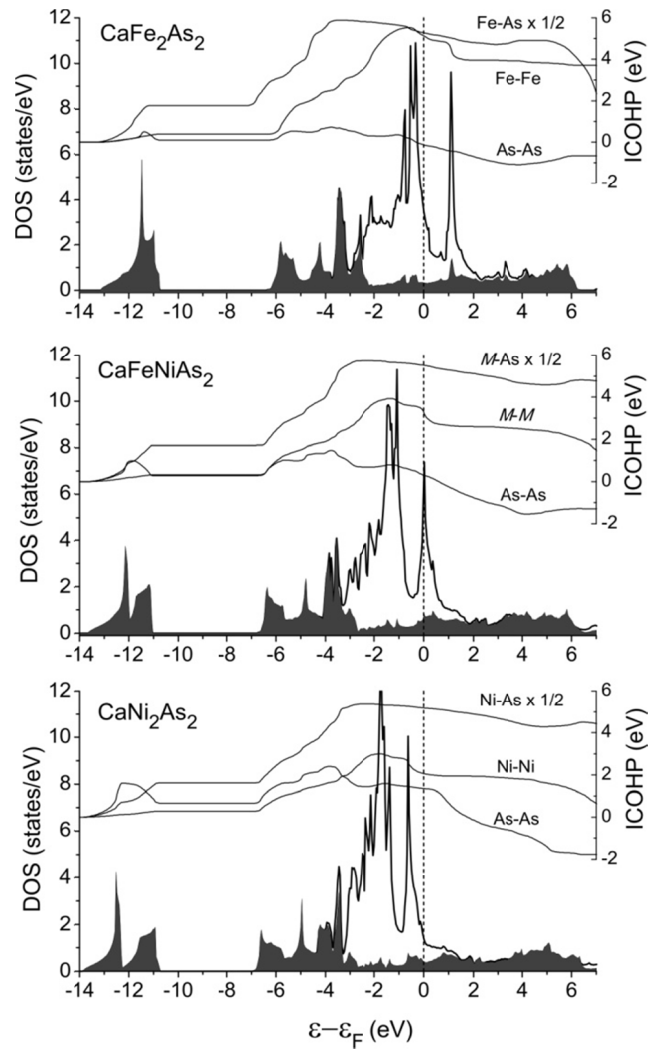

Fig. 5. Partial density of states (PDOS) of the metal 3$d$- (lines) and arsenic 4$s$/$p$ contributions (filled areas), together with the ICOHP curves of the $M$-As, $M$-$M$ and As-As bonds, respectively.

The ICOHP plots in Fig. 5 reflect the strengths of the bonding interactions. The $M$-As bonds are by far the strongest (~10 eV/cell), and remain almost unaffected by the VEC. Remarkably, the Fe-Fe bonds in CaFe$_2$As$_2$ (VEC = 28) are the second strongest (4.5 eV/cell), while the As-As bonding contribution is almost zero. When the VEC increases to 30 in CaFeNiAs$_2$, the $M$-$M$ bonds become significantly weaker (3 eV/cell, −33%). We point out, that this is only partially caused by filling of the antibonding states near $E_F$, because the $M$-$M$ ICOHP is inherently much smaller than in the case of iron as a result of the weaker overlap of the contracted Ni 3$d$-orbitals. The As-As bond energy is still very small and negligible at the As-As distance of 280 pm in CaFeNiAs$_2$. These trends continue in CaNi$_2$As$_2$, where the Ni-Ni bond energy is reduced to 2 eV/cell, that is less 50% of iron compound. The As-As distance is only 260 pm now, and the bonding energy contribution becomes significant, however their energy contribution is only ~1 eV/cell. The Fermi energy is between As-As bonding and antibonding states in all three compounds, and it is the bonding character of the bands that changes rather than the band occupation. The changes in the ICOHP bond energies with the valence electron count are compiled in Fig. 6. Note that the metal-metal-bonds still dominate over the As-As bonds even in CaNi$_2$As$_2$ with VEC =32.

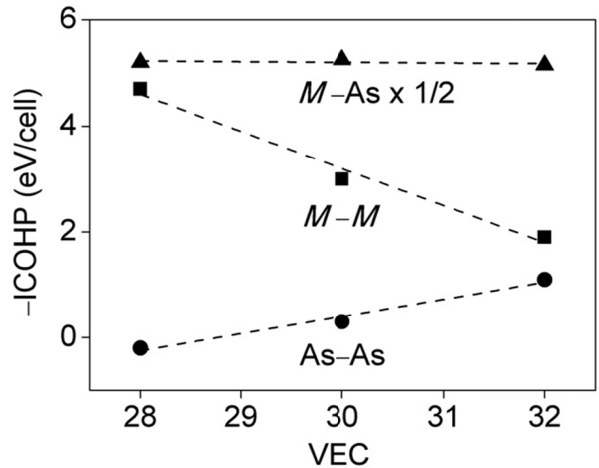

Fig. 6. Calculated ICOHP bond energies against the valence electron count ($M$ = Fe, Ni).

Our analysis reveals that the increasing electron count mainly affects the metal-metal bonds. Higher VEC leads to weaker *M-M* bonds, which causes longer *a,b* lattice parameters and flatter *M*As$_4$-tetrahedra. As a result, the interlayer distance becomes smaller and the interlayer bonds can be formed. Thus the formation of the As$_2$-dimers is not caused by a depopulation of As-As σ*-orbitals as anticipated earlier, but a consequence of the weakened *M-M* bonding in the *M*As$_{4/4}$-layer.

The magnetism observed in the solid solution Ca(Fe$_{1-x}$Ni$_x$)$_2$As$_2$ close to *x* = 0.5 is also understandable from the electronic structure. When the VEC is around 30 (CaFeNiAs$_2$), the Fermi energy coincides with the peak in the DOS (Fig. 5), and the Stoner criterion that favours a magnetic ground state is fulfilled. Magnetic ordering in CaCo$_2$As$_2$ [24] and CaCo$_2$P$_2$ [25] supports this argument. We also believe that the recently reported weak ferromagnetism in SrCo$_2$(Ge$_{0.5}$P$_{0.5}$)$_2$ (VEC = 29) is caused by this DOS-peak in the Co-3*d* states, and not by the intermediate *X-X* (*X* = Ge,P) bonding state as suggested in [22].

**Conclusion**

The structures of the solid solution Ca(Fe$_{1-x}$Ni$_x$)$_2$As$_2$ show the transition between the ThCr$_2$Si$_2$-type without As-As bonds between the layers (better referred to as the BaZn$_2$P$_2$-type [26, 27]) and the true ThCr$_2$Si$_2$-type structure with As$_2$-dimers. Our results indicate that the role of the homonuclear bonds between the layers of the ThCr$_2$Si$_2$-type structure is mostly overrated. The As 4*p*-orbitals spread over a large energy range, therefore any shifts of the Fermi level hardly affect their occupation. Thus the earlier interpretation of "making and breaking bonds" in the ThCr$_2$Si$_2$-type compounds does not correspond to the real situation. The electronic states at the Fermi level are clearly dominated by the transition metal 3*d*-band, in particular a large peak in the DOS coincides with the Fermi energy when the band filling is close to 30 electrons per formula unit. Even though this peak is antibonding with respect to the metal-metal bonds in the layers, the ICOHP analysis shows that its occupation has no strong effect in weakening the metal-metal bonds. Nevertheless it turned out that the metal-metal bonds inside the layers are the by far most affected. Indeed, the *M-M* bonding character of occupied states strongly decreases on going from iron to nickel, because the proceeding contraction of the 3*d*-orbitals reduces the overlap and weakens the *M-M* bonds. We note that 3*d*-orbital contraction had been also one argument of the Hoffmann model, but in the sense that this lowers the Fermi level and depopulates P-P $\sigma$*-orbitals, which is ruled out by our first-principle calculations.

Also the occurrence of the magnetic groundstate the ThCr$_2$Si$_2$-type compounds with 3*d*-transition metals and nearly 30 valence electrons in not connected to dimer making or breaking as suggested in a recent study [22]. Weak ferromagnetism develops near $x \approx 0.5$ in the solid solution Ca(Fe$_{1-x}$Ni$_x$)$_2$As$_2$, where the VEC is close to 30. This magnetism is a consequence of a DOS peak from flat 3*d*-bands, which coincides with the Fermi-level at 30 electrons per unit cell. In this case the Stoner criterion is fulfilled and ferromagnetism can emerge without significant contribution of the pnictide orbitals.

**Experimental Section**

*Synthesis and X-ray powder diffraction*

Polycrystalline samples were synthesized in alumina crucibles by mixing stoichiometric amounts of Ca (99.99 %), Fe (99.9 %), Ni (99.99 %) and As (99.999 %) in a glove box with purified Argon atmosphere. The crucibles were subsequently sealed in silica tubes under Argon atmosphere. The reaction mixtures were then heated to 773 K for 10 h, to 1033 K for 15 h and to 1273 K for 15 h before cooling to room temperature at a rate of 100 K/h. This first reaction step was followed by two annealing steps at 1173 K of which the second one was performed after pressing the homogenized sample into a pellet. The samples were characterized using X-ray powder diffraction with Cu-K$_{\alpha 1}$ and Co-K$_{\alpha 1}$ radiation respectively (HUBER G670 Guinier imaging plate diffractometer). Rietveld refinements were performed using the TOPAS program package [28].

*Magnetic measurements*

Magnetic susceptibility measurements were performed on a Quantum Design MPMS XL5 SQUID magnetometer which allowed for measurements with fields between -50 kOe and 50 kOe at temperatures between 1.8 K and 400 K.

*Electronic structure calculations*

Self-consistent DFT band structure calculations were performed using the LMTO-method in its scalar-relativistic version (program TB-LMTO-ASA)[29-31]. Reciprocal space integrations were performed with the tetrahedron method using 3013 irreducible *k*-points in the tetragonal Brillouin zone. The basis sets were Ca-4s/[4p]/3d, Fe(Ni)-4s/4p/3d and As-4s/4p/[4d]. Orbitals in brackets were downfolded. The COHP (Crystal orbital Hamilton population) [32, 33] method was used for the bond analysis. COHP gives the energy contributions of all electronic states for a selected bond. The values are negative for bonding and positive for antibonding interactions. With respect to the widely used COOP diagrams, we plot –COHP(E) to get positive values for bonding states.

*Acknowledgment*

This work has been supported financially by the German Research Foundation DFG within priority program SPP1458 under grant JO257/6.